\documentclass[12pt, a4paper]{article}

\usepackage[utf8]{inputenc} 
\usepackage{amsmath}        
\usepackage{amssymb}        
\usepackage{graphicx}       
\usepackage{geometry}       
\usepackage{hyperref}       
\usepackage{booktabs}       
\usepackage{float}          
\usepackage{sectsty}        
\usepackage{xcolor}         
\usepackage{titlesec}
\titleformat{\paragraph}[runin]{\bfseries}{\theparagraph}{1em}{}[.]
\hypersetup{
    colorlinks=true,
    linkcolor=blue,
    filecolor=magenta,
    urlcolor=cyan,
}

\title{\Large\textbf{\textcolor{black}{A Geomechanically-Informed Framework for Wellbore Trajectory Prediction: Integrating First-Principles Kinematics with a Rigorous Derivation of Gated Recurrent Networks}}}

\author{
Shubham Kumar\\
Department of Geology and Geophysics\\
IIT Kharagpur\\
shubhamkumarprasadsah@gmail.com
\and
Anshuman Sahoo\\
Department of Metallurgical and Materials Engineering\\
IIT Kharagpur\\
anshumanrkl05@gmail.com
}

\date{}

\begin{document}

\maketitle

\begin{abstract}
Precise wellbore trajectory prediction is a crucial task in subsurface engineering, dictated by nonlinear interactions among the drilling assembly and inhomogeneous geology. This study devises a mathematically well-posed framework for trajectory prediction that transitions beyond empirical modeling to a geomechanically-informed framework. The project makes use of Log ASCII Standard and wellbore deviation (DEV) measurements in 14 Gulfaks oil field wells, viewing petrophysical logs as surrogates to the mechanical properties of the rock that truly dictate drilling dynamics. An important result of the project is the formal derivation of wellbore kinematic models presenting them as proper numerical integration schemes. The heart of the predictive model is a Gated Recurrent Unit (GRU) network. The theoretical justification, often overlooked in practice-oriented works, details the mechanisms by which the network learns the dependencies over time. The methodology comprises a theoretically justified preprocessing of the data, i.e. invariant depth resampling, and sequence formation. The post-processing of the trajectories and the error analysis are carried out by way of MAE, RMSE and R2 methods. The results demonstrate that the GRU model, efficiently learns the implicit, nonlinear function of transformation from geology to directional shift, and succeeds in predicting the azimuth, inclination, and spatial location accurately. The present study can be regarded as the guideline for the construction of physically-based machine learning models for petroleum engineering, in which the network learns a functional description of the local Mechanical Earth Model, and can provide more accurate well planning and real-time geosteering operations.
\end{abstract}

\section{Introduction}

\subsection{The Physical Problem of Subsurface Navigation}
Directional drilling, the art and science of guiding a drill bit for thousands of meters through subterranean rock to intersect a targeted reservoir, is the foundation of contemporary energy production. The trajectory of a wellbore is not a geometric line but the physical expression of a dynamic, nonlinear equilibrium of forces \hyperlink{mitchell1995}{(Mitchell, 1995)}. These forces result from the mechanical coupling of the bottom hole assembly (BHA), the spinning drill bit, and adjacent rock formations, which frequently are anisotropic, fractured, and heterogeneous. The driller's means to control the trajectory are essential to maximize reservoir contact, to circumvent geologic hazards, and to make the asset commercially acceptable.

Traditional trajectory planning schemes generally rely on the extrapolation of geometry or reduced analytical models that cannot encompass the entire physics of the drilling process. These models cannot adequately predict how the wellbore would respond to unforeseen changes in the subsurface environment. The effectiveness of both pre-planed drilling and in real-time geosteering is inevitably limited by the ability to predict the wellbore path under the influence of the geology that it is about to enter.

\subsection{Geomechanical Drivers of Trajectory Deviation}
The deviation of the wellbore from the designed path is not random, but a deterministic reaction to the geomechanical environment. The main initiators of that deviation reside in the foundation of the law of rock mechanics and geologic structures. The overall understanding of that environment is generally summarized in the form of a Mechanical Earth Model (MEM), a 3D, multi-scale description of the properties of the rocks and in-situ stresses. The main geomechanical initiators are as follows:
    \paragraph{In-situ Stress Fields:} The subsurface is under a three-dimensional stress tensor, consisting of the vertical stress (due to the overburden weight) and the horizontal stresses. If the stresses are not equal (anisotropic), they produce preferred directions of failure and deformation of the rocks around the wellbore that can guide the drill bit.
    
    \paragraph{Mechanical Properties of the Rock:} The inherent properties of the rock govern how the rock reacts to the stress forces of the drill bit. Quantities like the Unconfined Compressive Strength, Young's Modulus, and Poisson's Ratio are most important.These properties are related most closely to petrophysical log measurements like porosity, permeability, and sonic velocity.
    \paragraph{Lithological Heterogeneity and Discontinuities:} The subsurface formations hardly comprise homogeneous sections. The path is greatly affected by changes in the lithology, that have varying mechanical properties. These interfaces, easily discerned by the gamma ray log and facies description, serve as the mechanical discontinuity that can generate sudden changes in the well path.

\subsection{Data-Driven Surrogate Modeling for a Complex System}
Explicitly simulating the entire coupled physics of bit-rock engagement, fluid flow, and drill string dynamics is a daunting task. The computational challenge thus provides the impetus for the creation of a data-driven surrogate model. A surrogate model does not try to numerically solve the governing partial differential equations of the system based on first principles. Rather, the surrogate learns the input-output character of the system from observation-based data.

Here, the Gated Recurrent Unit (GRU) neural network, an advanced extension of the Recurrent Neural Network (RNN), is leveraged not as an all-purpose, general-purpose "black-box" algorithm, but as a specially designed sequence processor especially well suited to the physics-based problem at hand. The sequential nature of drilling, where the current state depends functionally upon the formation history that has come before, translates directly to the recurrent structure of the GRU. The model learns the implicit, very nonlinear transform function that takes as input a sequence of geology measurements (the petrophysical logs) and produces as output the resulting kinematic wellbore state (inclination and azimuth).

Another main thesis of the current work is that a well-trained GRU does not simply learn to identify patterns; instead, it learns a functional, latent description of the local geomechanical environment. The internal hidden state vector of the network at any given depth can formally be understood as a compressed, low-dimensional representation of the geomechanically relevant steering properties. The model implicitly creates a log-space MEM of the local geomechanically active environment, an idea that offers an extremely useful physical interpretation of the network's intrinsic workings and prediction ability.

\section{Theoretical Background and Related Work}

\subsection{Mathematical Models of Wellbore Kinematics}
The trajectory of a wellbore is a continuous curve in three-dimensional Euclidean space, $\mathbb{R}^3$. We can represent this curve as a vector function $\mathbf{r}(s)$, parameterized by the measured depth $s$, which is the arc length along the wellbore from a reference point (e.g., the surface location).

\subsubsection{The Tangent Vector in Spherical Coordinates}
The local direction of the wellbore at any point $s$ is described by the unit tangent vector, $\mathbf{T}(s) = \frac{d\mathbf{r}}{ds}$. In directional drilling, this vector is conventionally defined by two angles: the inclination $I(s)$, which is the angle from the vertical axis (Z-axis), and the azimuth $A(s)$, which is the angle in the horizontal plane (X-Y plane) measured clockwise from the North direction (Y-axis). The components of the unit tangent vector are given by a standard spherical-to-Cartesian coordinate transformation:
\begin{equation}
\mathbf{T}(s) = \begin{pmatrix} \sin(I(s)) \cos(A(s)) \\ \sin(I(s)) \sin(A(s)) \\ \cos(I(s)) \end{pmatrix}
\end{equation}
Here, the coordinate system is defined with X as East, Y as North, and Z as True Vertical Depth (TVD) pointing downwards. Note that some conventions may swap X and Y or the direction of Z.

\subsubsection{The Average Angle Method as a Numerical Integration Scheme}
Given the tangent vector, the position of the wellbore can be found by integrating the differential equation $\frac{d\mathbf{r}}{ds} = \mathbf{T}(s)$. The total displacement vector from a point $s_i$ to $s_{i+1}$ is given by the definite integral:
\begin{equation}
\Delta \mathbf{r}_i = \mathbf{r}(s_{i+1}) - \mathbf{r}(s_i) = \int_{s_i}^{s_{i+1}} \mathbf{T}(s) \,ds
\end{equation}
In practice, survey data provides discrete measurements of inclination and azimuth, $(I_i, A_i)$, only at specific stations, $s_i$. Therefore, this integral must be approximated numerically. The Average Angle method is one such numerical scheme. It approximates the integral by assuming the tangent vector is constant over the interval $[s_i, s_{i+1}]$ and equal to the tangent vector evaluated at the average of the angles at the start and end of the interval.

Let $\Delta s_i = s_{i+1} - s_i$ be the step length along the measured depth. The average inclination and azimuth are defined as:
\begin{equation}
\bar{I}_i = \frac{I_i + I_{i+1}}{2}
\end{equation}
\begin{equation}
\bar{A}_i = \frac{A_i + A_{i+1}}{2}
\end{equation}
The integral is then approximated as:
\begin{equation}
\Delta \mathbf{r}_i \approx \Delta s_i \cdot \mathbf{T}(\bar{I}_i, \bar{A}_i) = \Delta s_i \begin{pmatrix} \sin(\bar{I}_i) \cos(\bar{A}_i) \\ \sin(\bar{I}_i) \sin(\bar{A}_i) \\ \cos(\bar{I}_i) \end{pmatrix}
\end{equation}
Writing out the components gives the displacement equations used in the source material:
\begin{equation}
\Delta X_i = \Delta s_i \cdot \sin\left(\frac{I_i + I_{i+1}}{2}\right) \cos\left(\frac{A_i + A_{i+1}}{2}\right)
\end{equation}
\begin{equation}
\Delta Y_i = \Delta s_i \cdot \sin\left(\frac{I_i + I_{i+1}}{2}\right) \sin\left(\frac{A_i + A_{i+1}}{2}\right)
\end{equation}
\begin{equation}
\Delta Z_i = \Delta s_i \cdot \cos\left(\frac{I_i + I_{i+1}}{2}\right)
\end{equation}
This derivation shows that the Average Angle method is mathematically equivalent to applying the midpoint rule for numerical integration to the vector ordinary differential equation governing the trajectory. The total trajectory is then found by the cumulative sum of these incremental displacement vectors: $\mathbf{r}_{i+1} = \mathbf{r}_i + \Delta \mathbf{r}_i$.

\subsubsection{The Minimum Curvature Method}
A more sophisticated and widely adopted approach is the Minimum Curvature method \hyperlink{sawaryn2005}{(Sawaryn and Thorogood, 2005)}. This method models the wellbore segment between two survey stations, $\mathbf{r}_i$ and $\mathbf{r}_{i+1}$. This assumption is physically motivated by the idea that a drill string, under tension and compression, tends to form smooth curves that minimize bending energy. The path is assumed to lie in a plane whose orientation is determined by the tangent vectors $\mathbf{T}_i$ and $\mathbf{T}_{i+1}$ at the endpoints.

The central angle of this arc, known as the dogleg angle $\beta_i$, is the angle between the two tangent vectors and can be found from their dot product:
\begin{equation}
\cos(\beta_i) = \mathbf{T}_i \cdot \mathbf{T}_{i+1}
\end{equation}
Substituting the components of the tangent vectors yields:
\begin{equation}
\cos(\beta_i) = \sin(I_i)\sin(I_{i+1})\cos(A_i)\cos(A_{i+1}) + \sin(I_i)\sin(I_{i+1})\sin(A_i)\sin(A_{i+1}) + \cos(I_i)\cos(I_{i+1})
\end{equation}
Using trigonometric identities, this simplifies to:
\begin{equation}
\cos(\beta_i) = \cos(I_{i+1} - I_i) - \sin(I_i)\sin(I_{i+1})(1 - \cos(A_{i+1} - A_i))
\end{equation}
The displacement vector is then calculated by multiplying the average of the tangent vectors by the measured depth step and a Ratio Factor (RF) that corrects for the curvature:
\begin{equation}
\Delta \mathbf{r}_i = \frac{\Delta s_i}{2} (\mathbf{T}_i + \mathbf{T}_{i+1}) \cdot \text{RF}_i
\end{equation}
where the Ratio Factor is given by:
\begin{equation}
\text{RF}_i = \frac{2}{\beta_i} \tan\left(\frac{\beta_i}{2}\right)
\end{equation}
When the dogleg angle $\beta_i$ is small, $\tan(\beta_i/2) \approx \beta_i/2$, and the RF approaches 1, causing the Minimum Curvature method to converge to the Balanced Tangential method (a close relative of the Average Angle method). The Minimum Curvature method is generally considered the industry standard for its accuracy in representing smoothly curving wellbores.

\subsubsection{Dogleg Severity as a Measure of Local Curvature}
Dogleg Severity (DLS) is a critical parameter in drilling engineering that quantifies the total curvature of the wellbore over a given interval. High DLS can induce excessive stress on the drill pipe and casing, leading to fatigue and failure. Mathematically, DLS is an approximation of the geometric curvature, $\kappa$, of the path.

Curvature is formally defined as the magnitude of the rate of change of the unit tangent vector with respect to arc length:
\begin{equation}
\kappa(s) = \left\| \frac{d\mathbf{T}}{ds} \right\|
\end{equation}
For a discrete segment between $s_i$ and $s_{i+1}$, we can approximate this derivative using a finite difference:
\begin{equation}
\kappa \approx \left\| \frac{\mathbf{T}_{i+1} - \mathbf{T}_i}{\Delta s_i} \right\| = \frac{1}{\Delta s_i} \sqrt{(\mathbf{T}_{i+1} - \mathbf{T}_i) \cdot (\mathbf{T}_{i+1} - \mathbf{T}_i)}
\end{equation}
Expanding the dot product and using the fact that $\|\mathbf{T}_i\| = \|\mathbf{T}_{i+1}\| = 1$, we get:
\begin{equation}
\kappa \approx \frac{1}{\Delta s_i} \sqrt{2 - 2(\mathbf{T}_i \cdot \mathbf{T}_{i+1})} = \frac{1}{\Delta s_i} \sqrt{2 - 2\cos(\beta_i)}
\end{equation}
Using the half-angle identity $\sin(\beta_i/2) = \sqrt{(1-\cos(\beta_i))/2}$, this simplifies to:
\begin{equation}
\kappa \approx \frac{2 \sin(\beta_i/2)}{\Delta s_i}
\end{equation}
For small angles, $\sin(\beta_i/2) \approx \beta_i/2$, so $\kappa \approx \beta_i / \Delta s_i$. The dogleg angle $\beta_i$ is precisely the angle whose cosine was derived in the previous section. Therefore, the DLS formula presented in the source material is a direct calculation of this angle, normalized to a standard length (e.g., 100 ft or 30 m):
\begin{equation}
\text{DLS} = \frac{C}{\Delta s_i} \arccos[\cos(I_{i+1} - I_i) - \sin(I_i)\sin(I_{i+1})(1 - \cos(A_{i+1} - A_i))]
\end{equation}
where $C$ is the normalization constant (e.g., 100 ft). This derivation firmly grounds DLS in the differential geometry of curves, identifying it as the average rate of change of direction over the measured interval.

\subsection{Geomechanical Influences on Drilling Dynamics}
The models we presented above give us the vocabulary to talk about the path of a wellbore, but they cannot tell us \textit{why} the path is not planar. The reason is the coupling of the drilling assembly with the geomechanic environment.

\subsubsection{From Petrophysical Logs to Mechanical Properties}
Petrophysical well logs, while no actual measurement of the strength of the rock, are good surrogates of the mechanics governing drillability and wellbore stability. The properties used as inputs in the present study (Gamma Ray, Porosity, Permeability, Fluvial Facies, Net-to-Gross) are all common inputs in the industry for the build-up of MEMs.

    \paragraph{Gamma Ray (GAMMA):} This log assesses natural radioactivity and is the most important indicator of lithology, separating clean sandstones (low GAMMA) and shales (high GAMMA). The shales are generally weaker and more plastic than sandstones, resulting in varying drilling responses.
    \paragraph{Porosity (POROSITY):} Void space in a rock. Higher porosity generally indicates lower strength of the rock and lower Young's Modulus (rigidity).
    \paragraph{Permeability (PERM):} Measure of how much a rock can let in or pass by fluids. Though not a direct indicator of strength, it has frequently been correlated with porosity and grain diameter, both of which generally reduce strength. Furthermore, very high permeability can affect the pressure balance at the bit face, affecting stability.
    \paragraph{Fluvial Facies (FLUVIALFACIES) \& Net-to-Gross (NETGROSS):} These are interpreted geology attributes that give a higher-level background. Facies categories give an indication of the depositational environment and are accompanied by particular lithologies as well as rock textures. Net-to-Gross expresses the quantity of reservoir quality rock, that often differs in mechanics from non-reservoir rock.

\begin{figure}[H]
\centering
\includegraphics[width=0.9\textwidth]{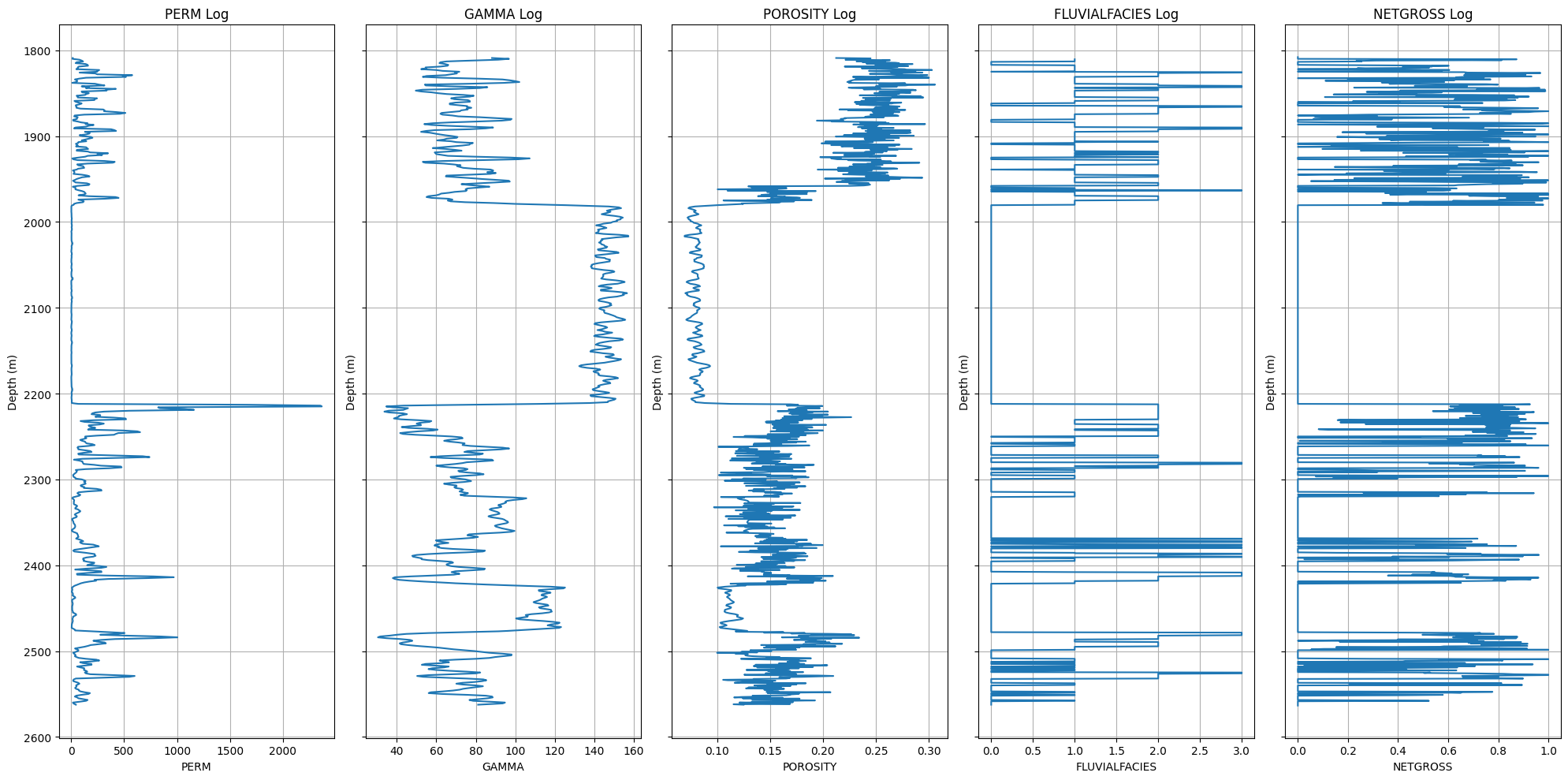}
\caption{Sample Log Data}
\label{fig:sample_logs}
\end{figure}

\subsection{Evolution of Recurrent Models for Sequential Data}

\subsubsection{The Simple Recurrent Neural Network (RNN)}
The foundational model for processing sequential data is the Simple RNN \hyperlink{rumelhart1986}{(Rumelhart et al., 1986)}. At each time step $t$, an RNN takes an input vector $\mathbf{x}_t$ and the hidden state from the previous time step, $\mathbf{h}_{t-1}$, to compute the new hidden state $\mathbf{h}_t$:
\begin{equation}
\mathbf{h}_t = \sigma_h (\mathbf{W}_{xh} \mathbf{x}_t + \mathbf{W}_{hh} \mathbf{h}_{t-1} + \mathbf{b}_h)
\end{equation}
The output $\mathbf{y}_t$ is then typically a function of the hidden state:
\begin{equation}
\mathbf{y}_t = \sigma_y (\mathbf{W}_{hy} \mathbf{h}_t + \mathbf{b}_y)
\end{equation}
where $\mathbf{W}_{xh}, \mathbf{W}_{hh}, \mathbf{W}_{hy}$ are shared weight matrices, $\mathbf{b}_h, \mathbf{b}_y$ are bias vectors, and $\sigma_h, \sigma_y$ are activation functions (e.g., hyperbolic tangent). The recurrence relation for $\mathbf{h}_t$ allows the network to maintain a "memory" of past inputs, encoded in the hidden state vector.

\subsubsection{The Vanishing and Exploding Gradient Problem}
Recurrent Neural Networks remember from previous jobs, which is a plus point. But teaching them is not that easy because of the vanishing and exploding gradient problems \hyperlink{hochreiter1997}{(Hochreiter and Schmidhuber, 1997)}. What makes them hard to train is where BPTT calculates gradient by unrolling the network through time layers.

Consider the gradient of the loss at time step $T$ in the hidden state at a much earlier time step $t \ll T$. The chain rule dictates that it's a product of Jacobian matrices:
\begin{equation}
\frac{\partial \mathcal{L}_T}{\partial \mathbf{h}_t} = \frac{\partial \mathcal{L}_T}{\partial \mathbf{h}_T} \frac{\partial \mathbf{h}_T}{\partial \mathbf{h}_{T-1}} \frac{\partial \mathbf{h}_{T-1}}{\partial \mathbf{h}_{T-2}} \cdots \frac{\partial \mathbf{h}_{t+1}}{\partial \mathbf{h}_t}
\end{equation}
Each Jacobian term is $\frac{\partial \mathbf{h}_k}{\partial \mathbf{h}_{k-1}} = \text{diag}(\sigma_h'(\cdot)) \mathbf{W}_{hh}$. Multiplication of the matrix $\mathbf{W}_{hh}$ over and over in several time steps regulates the output of the resulting product. If the maximum singular value of the matrix $\mathbf{W}_{hh}$ exceeds 1, the gradient norm increases too rapidly, so the gradients explode, and the learning is unstable. However, whenever the maximum singular value is less than 1, the norm decreases too rapidly, so the gradients vanish, and the network cannot learn long-range dependencies anymore.

\subsubsection{Architectural Solutions: LSTM and GRU}
In order to break through this basic limitation, more advanced recurrent architectures became available. The Long Short-Term Memory (LSTM) network was an originally innovative solution \hyperlink{hochreiter1997}{(Hochreiter and Schmidhuber, 1997)}. LSTMs offer a distinct "cell state" vector that is an information highway, where information propagates through time with very few linear interactions.

The Gated Recurrent Unit (GRU), is a subsequent, less computationally intensive replacement for the LSTM. It concatenates the forget and the input gate into a shared "update gate" and merges the cell state and the hidden state \hyperlink{cho2014}{(Cho et al., 2014)}. It has a "reset gate" that decides how much of the previous hidden state to contribute in the calculation of the new candidate hidden state.

\section{Mathematical and Computational Methodology}
This section provides a complete, first-principles derivation of the entire modeling pipeline.

\begin{table}[H]
\centering
\caption{Table of Notations}
\label{tab:notations}
\begin{tabular}{@{}lll@{}}
\toprule
Symbol & Description & Dimensions \\ \midrule
$s$ & Measured Depth (arc length) & Scalar \\
$\mathbf{r}(s)$ & Position vector of the wellbore & $3 \times 1$ \\
$I, A$ & Inclination and Azimuth angles & Scalar \\
$\mathbf{T}$ & Unit tangent vector & $3 \times 1$ \\
$t$ & Discrete time step (depth index) & Integer \\
$\mathbf{x}_t$ & Input feature vector at step $t$ & $d_x \times 1$ \\
$\mathbf{y}_t$ & Target vector (e.g., $[I_t, A_t]^T$) & $d_y \times 1$ \\
$\hat{\mathbf{y}}_t$ & Predicted target vector & $d_y \times 1$ \\
$\mathbf{h}_t$ & GRU hidden state vector at step $t$ & $d_h \times 1$ \\
$\mathbf{z}_t$ & GRU update gate vector & $d_h \times 1$ \\
$\mathbf{r}_t$ & GRU reset gate vector & $d_h \times 1$ \\
$\tilde{\mathbf{h}}_t$ & GRU candidate hidden state vector & $d_h \times 1$ \\
$\mathbf{W}_z, \mathbf{W}_r, \mathbf{W}_h$ & Input-to-hidden weight matrices & $d_h \times d_x$ \\
$\mathbf{U}_z, \mathbf{U}_r, \mathbf{U}_h$ & Hidden-to-hidden weight matrices & $d_h \times d_h$ \\
$\mathbf{b}_z, \mathbf{b}_r, \mathbf{b}_h$ & Bias vectors & $d_h \times 1$ \\
$\sigma(\cdot)$ & Sigmoid activation function & Element-wise \\
$\tanh(\cdot)$ & Hyperbolic tangent activation function & Element-wise \\
$\odot$ & Element-wise (Hadamard) product & - \\
$\mathcal{L}$ & Loss function & Scalar \\
$\alpha$ & Learning rate & Scalar \\
$\mathbf{g}_t$ & Gradient of the loss w.r.t. parameters & Varies \\
$\mathbf{m}_t, \mathbf{v}_t$ & Adam first and second moment estimates & Varies \\
$\beta_1, \beta_2$ & Adam exponential decay rates & Scalar \\
$d_x, d_y, d_h$ & Dimensionality of input, output, hidden state & Integer \\
$w$ & Sequence window size & Integer \\ \bottomrule
\end{tabular}
\end{table}

\subsection{Formal Problem Definition}
The goal of this research is to create a predictive model for wellbore trajectory. The physical system is modeled by a sequence of discrete data points sampled along the wellbore. The input data is a multivariate sequence of petrophysical measurements, $\{\mathbf{x}_t\}_{t=1}^N$, where $\mathbf{x}_t \in \mathbb{R}^{d_x}$ is the $d_x=5$-dimensional vector of features (GAMMA, POROSITY, PERM, FLUVIALFACIES, NETGROSS) at the $t$-th depth step. The target vectors are the following changes in inclination and azimuth, which specify the future path.

The learning problem is posed as a sequence-to-vector forecasting task. For a given input feature history window of size $w$, the model needs to forecast the target vector $\mathbf{y}_t = [I_t, A_t]^T \in \mathbb{R}^{d_y}$ for the current time step. The model, represented by the function $f_{\boldsymbol{\theta}}$ with parameters $\boldsymbol{\theta}$, seeks to learn the following mapping:
\begin{equation}
\hat{\mathbf{y}}_t = f_{\boldsymbol{\theta}}(\mathbf{x}_{t-w+1}, \mathbf{x}_{t-w+2}, \dots, \mathbf{x}_t)
\end{equation}
The objective of the learning procedure is to determine the best set of parameters $\boldsymbol{\theta}^*$ that minimizes the loss function $\mathcal{L}(\mathbf{y}_t, \hat{\mathbf{y}}_t)$ over all training sequences available.

\subsection{Data Preprocessing as Mathematical Transformations}
Raw well log data requires several transformation steps to be suitable for input into a neural network. Each step has a clear mathematical basis and rationale.

\subsubsection{Feature Space Normalization}
Well log measurements have quite distinct numerical ranges and physical units. Features with larger magnitudes must be prevented from dominating the learning process and inducing unstable gradient descent, and the features should be scaled to the same range. For this work, Min-Max scaling, a linear transformation that scales all the features to the interval [0,1], is applied. For some feature vector $\mathbf{x}$, the Min-Max transformation operator $T_{mm}$ is defined by:
\begin{equation}
\mathbf{x}_{\text{norm}} = T_{mm}(\mathbf{x}) = \frac{\mathbf{x} - \mathbf{x}_{\min}}{\mathbf{x}_{\max} - \mathbf{x}_{\min}}
\end{equation}
where $ \mathbf{x}_{\min} $ and $ \mathbf{x}_{\max} $ are the minimum and maximum of the feature in the training data, and division is element-wise.

Another is Z-score standardization, which rescales features to mean 0 and standard deviation 1:
\begin{equation}
\mathbf{x}_{\text{std}} = T_z(\mathbf{x}) = \frac{\mathbf{x} - \boldsymbol{\mu}}{ \boldsymbol{\sigma}}
\end{equation}
where $\boldsymbol{\mu}$ and $\boldsymbol{\sigma}$ are mean and standard deviation vectors. Min-Max scaling was used here since activation functions used within the GRU gates (sigmoid and tanh) are saturating ones. Scaling the input features to [0,1] by Min-Max scaling ensures that the inputs to the activation functions lie in their most sensitive, i.e., non-saturated regions of the function, which is sure to improve learning.

\subsubsection{Temporal Discretization and Interpolation}
Well log data is often recorded at uneven depth intervals. RNN architectures like the GRU work with sequences that have fixed, uniform time steps. So, the data needs to be resampled onto a uniform depth grid. This process involves two steps: defining the grid and estimating values at the new grid points.

A uniform grid of measured depths $\{s_k\}_{k=1}^{M}$ is created with a constant step size $\Delta s = 0.5$ m. This starts from the minimum depth and ends at the maximum depth in the well. For any grid point $s_k$ that does not match an original measurement, its feature vector $\mathbf{x}(s_k)$ must be estimated from the nearby original data points. This study uses linear interpolation.

Given two consecutive original data points $(s_i, \mathbf{x}_i)$ and $(s_{i+1}, \mathbf{x}_{i+1})$, the interpolated value $\mathbf{x}(s_k)$ for any $s_k \in [s_i, s_{i+1}]$ is calculated using the formula:
\begin{equation}
\mathbf{x}(s_k) = \mathbf{x}_i + (\mathbf{x}_{i+1} - \mathbf{x}_i) \frac{s_k - s_i}{s_{i+1} - s_i}
\end{equation}
Choosing this interpolation method is not just a matter of convenience. More complex methods, like cubic spline interpolation, can create smoother series. However, they also risk adding unrealistic features. Geological formations can show sharp, abrupt changes at boundaries (e.g., faults, unconformities). A spline interpolator, which tries to keep derivative continuity, can "overshoot" or "ring" around these sharp transitions, causing oscillations that do not exist in the actual geology. A strong learning model like a GRU might mistakenly recognize these induced oscillations as real high-frequency geological signals, leading it to learn false correlations and reducing its ability to adapt to new data. Linear interpolation assumes a simpler, piecewise-linear model of the subsurface. This is a more cautious and reliable choice that is less likely to create misleading artifacts.

\subsection{Gated Recurrent Unit (GRU) Network: A First-Principles Formulation}

\subsubsection{Forward Propagation Dynamics}
The GRU cell computes its new hidden state $\mathbf{h}_t$ from the current input $\mathbf{x}_t$ and previous hidden state $\mathbf{h}_{t-1}$ using a sequence of gating mechanisms. The following equations illustrate the operations of a single GRU cell at time step $t$.

\paragraph{Reset Gate ($\mathbf{r}_t$):} The reset gate controls how much of the previous information (from $\mathbf{h}_{t-1}$) is to be discarded while calculating the new candidate hidden state. It is computed as:
    \begin{equation}
    \mathbf{r}_t = \sigma(\mathbf{W}_r \mathbf{x}_t + \mathbf{U}_r \mathbf{h}_{t-1} + \mathbf{b}_r)
    \end{equation}
The sigmoid function $\sigma(z) = (1 + e^{-z})^{-1}$ confines the output to the $(0, 1)$ range. Near 0 values make the network "forget" the past, and near 1 values make it "remember".

\paragraph{Update Gate ($\mathbf{z}_t$):} The update gate determines the proportion of the new candidate state to be incorporated in order to update the hidden state, and in contrast, how much of the previous hidden state should be retained.
\begin{equation}
\mathbf{z}_t = \sigma(\mathbf{W}_z \mathbf{x}_t + \mathbf{U}_z \mathbf{h}_{t-1} + \mathbf{b}_z)
\end{equation}

\paragraph{Candidate Hidden State ($\tilde{\mathbf{h}}_t$):} This is a proposal for the new hidden state. It is calculated in a similar manner to a basic RNN's hidden state, but with the effect of the old hidden state weighted by the reset gate.
\begin{equation}
\tilde{\mathbf{h}}_t = \tanh(\mathbf{W}_h \mathbf{x}_t + \mathbf{U}_h (\mathbf{r}_t \odot \mathbf{h}_{t-1}) + \mathbf{b}_h)
\end{equation}
The element-wise multiplication $\mathbf{r}_t \odot \mathbf{h}_{t-1}$ enables the reset gate to selectively zero out elements of the last hidden state, effectively regulating its effect on the candidate state. The hyperbolic tangent function $\tanh(z) = \frac{e^z - e^{-z}}{e^z + e^{-z}}$ scales the output to $(-1, 1)$.

\paragraph{Hidden State Update ($\mathbf{h}_t$):} The final hidden state is a convex combination of the previous hidden state $\mathbf{h}_{t-1}$ and the candidate state $\tilde{\mathbf{h}}_t$, with the update gate $\mathbf{z}_t$ controlling the mixture.
    \begin{equation}
    \mathbf{h}_t = (1 - \mathbf{z}_t) \odot \mathbf{h}_{t-1} + \mathbf{z}_t \odot \tilde{\mathbf{h}}_t
    \end{equation}
This additive form is essential for alleviating the vanishing gradient problem, as it allows for a straight pass of gradients through time, just like the cell state in an LSTM.

\begin{figure}[H]
\centering
\includegraphics[width=0.9\textwidth]{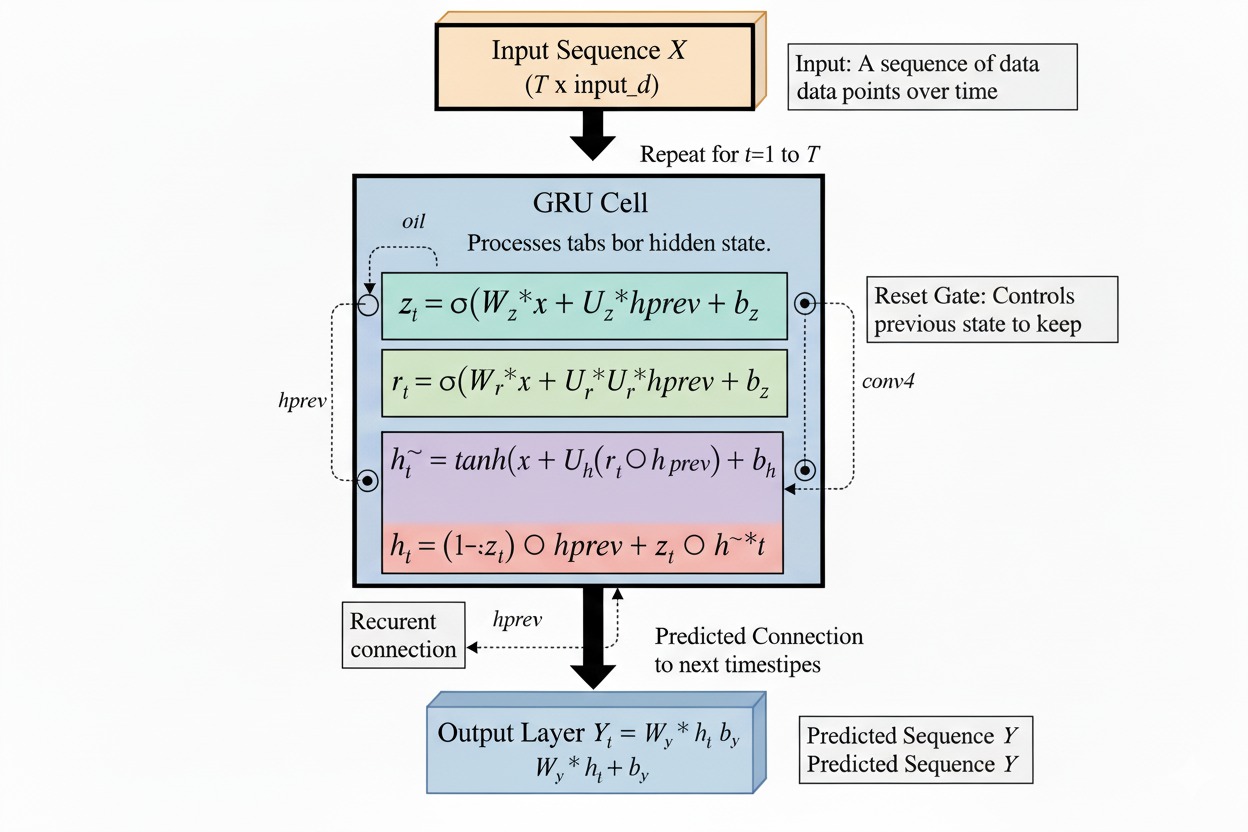}
\caption{Architecture of the GRU-based Trajectory Prediction Model}
\label{fig:architecture}
\end{figure}

\subsubsection{The Backpropagation Through Time (BPTT) Algorithm for GRUs}
Training the GRU requires computing the gradient of a total loss function $\mathcal{L}$ with respect to all model parameters $\boldsymbol{\theta} = \{\mathbf{W}_z, \mathbf{U}_z, \mathbf{b}_z, \dots\}$. The loss is typically summed over a sequence of length $T$: $\mathcal{L} = \sum_{t=1}^T \mathcal{L}_t$, where $\mathcal{L}_t$ is the loss at time step $t$. BPTT is an application of the chain rule to the unrolled computational graph of the RNN.

Let $\delta_t^h = \frac{\partial \mathcal{L}}{\partial \mathbf{h}_t}$ be the gradient of the total loss with respect to the hidden state at time step $t$. This represents the "error" signal that needs to be propagated backward. The gradient at time step $t$ depends on the gradient at $t+1$ and the gradient from the output at time step $t$.
\begin{equation}
\delta_t^h = \frac{\partial \mathcal{L}}{\partial \mathbf{h}_t} = \left(\frac{\partial \mathbf{h}_{t+1}}{\partial \mathbf{h}_t}\right)^T \delta_{t+1}^h + \left(\frac{\partial \mathcal{L}_t}{\partial \mathbf{h}_t}\right)^T
\end{equation}
The core of BPTT is to compute the Jacobian matrix $\frac{\partial \mathbf{h}_{t+1}}{\partial \mathbf{h}_t}$ and use this recurrence to propagate the error signal from $\delta_T^h$ back to $\delta_0^h$.

Let's derive the components of this Jacobian for the GRU. We start from the hidden state update equation: $\mathbf{h}_{t+1} = (1 - \mathbf{z}_{t+1}) \odot \mathbf{h}_t + \mathbf{z}_{t+1} \odot \tilde{\mathbf{h}}_{t+1}$.
Applying the multivariate chain rule:
\begin{equation}
\frac{\partial \mathbf{h}_{t+1}}{\partial \mathbf{h}_t} = \frac{\partial}{\partial \mathbf{h}_t} \left[ (1 - \mathbf{z}_{t+1}) \odot \mathbf{h}_t + \mathbf{z}_{t+1} \odot \tilde{\mathbf{h}}_{t+1} \right]
\end{equation}
This expands into several terms:
\begin{equation}
\frac{\partial \mathbf{h}_{t+1}}{\partial \mathbf{h}_t} = \text{diag}(1 - \mathbf{z}_{t+1}) - \text{diag}(\mathbf{h}_t) \frac{\partial \mathbf{z}_{t+1}}{\partial \mathbf{h}_t} + \text{diag}(\tilde{\mathbf{h}}_{t+1}) \frac{\partial \mathbf{z}_{t+1}}{\partial \mathbf{h}_t} + \text{diag}(\mathbf{z}_{t+1}) \frac{\partial \tilde{\mathbf{h}}_{t+1}}{\partial \mathbf{h}_t}
\end{equation}
Now we need the derivatives of the gates and the candidate state with respect to $\mathbf{h}_t$:
\begin{itemize}
    \item $\frac{\partial \mathbf{z}_{t+1}}{\partial \mathbf{h}_t} = \text{diag}(\sigma'(\cdot)) \mathbf{U}_z$
    \item $\frac{\partial \mathbf{r}_{t+1}}{\partial \mathbf{h}_t} = \text{diag}(\sigma'(\cdot)) \mathbf{U}_r$
    \item $\frac{\partial \tilde{\mathbf{h}}_{t+1}}{\partial \mathbf{h}_t} = \text{diag}(\tanh'(\cdot)) \left( \mathbf{U}_h \text{diag}(\mathbf{r}_{t+1}) + \mathbf{U}_h \text{diag}(\mathbf{h}_t) \frac{\partial \mathbf{r}_{t+1}}{\partial \mathbf{h}_t} \right)$
\end{itemize}
Substituting these back gives the full, complex expression for the state-to-state Jacobian. Once the error signals $\delta_t^h$ are computed for all $t=T, \dots, 1$, the gradients for the parameters can be found. For example, let's derive the gradient for $\mathbf{U}_z$:
\begin{equation}
\frac{\partial \mathcal{L}}{\partial \mathbf{U}_z} = \sum_{t=1}^T \frac{\partial \mathcal{L}}{\partial \mathbf{h}_t} \frac{\partial \mathbf{h}_t}{\partial \mathbf{z}_t} \frac{\partial \mathbf{z}_t}{\partial \mathbf{U}_z}
\end{equation}
Let $\mathbf{a}_z(t) = \mathbf{W}_z \mathbf{x}_t + \mathbf{U}_z \mathbf{h}_{t-1} + \mathbf{b}_z$. The error signal for the update gate's pre-activation is $\delta_t^{a_z} = \frac{\partial \mathcal{L}}{\partial \mathbf{a}_z(t)}$.
\begin{equation}
\delta_t^{a_z} = \left(\frac{\partial \mathbf{h}_t}{\partial \mathbf{z}_t} \frac{\partial \mathbf{z}_t}{\partial \mathbf{a}_z(t)}\right)^T \delta_t^h = \text{diag}(\sigma'(\mathbf{a}_z(t))) \text{diag}(\tilde{\mathbf{h}}_t - \mathbf{h}_{t-1}) \delta_t^h
\end{equation}
Then, the gradient for $\mathbf{U}_z$ is the sum of outer products over time:
\begin{equation}
\frac{\partial \mathcal{L}}{\partial \mathbf{U}_z} = \sum_{t=1}^T \delta_t^{a_z} (\mathbf{h}_{t-1})^T
\end{equation}
Similar derivations yield the gradients for all other weight matrices and bias vectors. This detailed process, while mathematically intensive, is what enables the network to learn the complex temporal patterns in the data by iteratively adjusting its parameters to minimize the prediction error.

\subsection{Parameter Optimization via Adaptive Moment Estimation (Adam)}
The computed gradients through BPTT are applied to update the model parameters. The Adam optimizer \hyperlink{kingma2015}{(Kingma and Ba, 2015)}, named after its creators, is a sophisticated stochastic gradient descent algorithm that has become a de facto standard for training deep neural networks. It effectively combines two other widely used extensions of SGD: AdaGrad, which keeps track of a per-parameter learning rate, and RMSProp, which uses per-parameter adaptive learning rates as well.

Adam calculates adaptive learning rates for all parameters using first and second moment estimates of gradients. Let $\boldsymbol{\theta}_t$ be the parameters at iteration $t$ and $\mathbf{g}_t = \nabla_{\boldsymbol{\theta}} \mathcal{L}_t$ be the gradient. The update rules are as follows:

    \paragraph{Update biased first moment estimate (momentum):} An exponentially weighted average of previous gradients.
    \begin{equation}
\mathbf{m}_t = \beta_1 \mathbf{m}_{t-1} + (1 - \beta_1) \mathbf{g}_t
    \end{equation}
    The hyperparameter $\beta_1$ (typically $\sim 0.9$) controls the decay rate. $\mathbf{m}_t$ acts as a momentum term, helping to accelerate descent in consistent directions and dampen oscillations.

    \paragraph{Update biased second moment estimate (adaptive learning rate):} An exponentially decaying average of past squared gradients.
\begin{equation}
    \mathbf{v}_t = \beta_2 \mathbf{v}_{t-1} + (1 - \beta_2) \mathbf{g}_t^2
    \end{equation}
    $\beta_2$ (usually $\sim 0.999$) adjusts the decay rate.
    $\mathbf{v}_t$ approximates the variance of gradients.

\paragraph{Calculate bias-corrected moment estimations:} The moment estimations $\mathbf{m}_t$ and $\mathbf{v}_t$ are initialized with vectors of zeroes and are therefore biased towards zero, particularly at the early stage of training. Adam corrects for such bias:
    \begin{equation}
    \hat{\mathbf{m}}_t = \frac{\mathbf{m}_t}{1 - \beta_1^t}
    \end{equation}
    \begin{equation}
    \hat{\mathbf{v}}_t = \frac{\mathbf{v}_t}{1 - \beta_2^t}
\end{equation}

    \paragraph{Update parameters:} This last update of the parameters is done with the bias-corrected estimates.
    \begin{equation}
    \boldsymbol{\theta}_{t+1} = \boldsymbol{\theta}_t - \alpha \frac{\hat{\mathbf{m}}_t}{\sqrt{\hat{\mathbf{v}}_t} + \epsilon}
    \end{equation}
In this, $\alpha$ is the learning rate on a global scale, and $\epsilon$ (e.g., $10^{-8}$) is a small constant to ensure numerical stability. The expression $\sqrt{\hat{\mathbf{v}}_t}$ serves to normalize the gradient efficiently, implementing per-parameter adaptive learning rate. Parameters with higher or more frequent gradients get smaller updates, whereas parameters with lower gradients get higher updates, resulting in faster convergence and stable convergence.

\section{Experimental Protocol}

\subsection{Dataset as a Discretized Physical System}
The paper uses a data set from the Gulfaks oil field, amounting to 14 individual wellbores' worth of data. The data set is treated as a sequence of discrete samples from a continuous, spatially varying geological volume. Each well provides a 1D transect across this 3D volume, recording both the geological properties encountered (LAS files) and the physical path followed (DEV files). The union of these two datasets by a depth-based matching procedure gives a resulting dataset in which the depth point of each feature vector has a corresponding directional state. The key characteristics of the processed dataset used for training and evaluation are summarized in Table \ref{tab:dataset}.

\begin{table}[H]
\centering
\caption{Dataset Characteristics}
\label{tab:dataset}
\begin{tabular}{@{}ll@{}}
\toprule
Parameter & Value \\ \midrule
Number of Wells & 14 \\
Feature Dimensions ($d_x$) & 5 (GAMMA, POROSITY, PERM, FLUVIALFACIES, NETGROSS) \\
Target Dimensions ($d_y$) & 2 (Inclination (INCL), Azimuth (AZIM)) \\
Resampling Interval ($\Delta s$) & 0.5 m \\
Sequence Window Size ($w$) & 50 \\
Sequence Stride & 10 \\
Total Sequences Generated & 1,594 \\ \bottomrule
\end{tabular}
\end{table}

\subsection{Model Configuration and Hyperparameter Rationale}
The GRU network design was intended to use a series of hyperparameters chosen so that model capacity and computational expense would be equated. The configuration is described in Table \ref{tab:hyperparams}. The reason the window width of 50 measurements, i.e., a 25-meter interval ($50 \times 0.5$ m), is chosen is geological. This range of thickness is typically sufficient to capture the average thickness of sedimentary elements, such as crevasse splays or fluvial channels, delineated in the fluvial facies log. This allows the model to observe a complete geological unit before prediction. The 64 hidden state dimension provides the model with ample capacity to learn a compact rich representation of such a 25-meter geologic terrain without being excessively susceptible to overfitting.

\begin{table}[H]
\centering
\caption{GRU Model Hyperparameters}
\label{tab:hyperparams}
\begin{tabular}{@{}ll@{}}
\toprule
Hyperparameter & Value \\ \midrule
Hidden Units ($d_h$) & 64 \\
Learning Rate ($\alpha$) & 0.001 \\
Batch Size & 64 \\
Training Epochs & 100 \\
Early Stopping Patience & 10 \\
Validation Split & 20\% \\ \bottomrule
\end{tabular}
\end{table}

\subsection{Training Protocol and Regularization}
The model was trained with a typical supervised learning procedure with some best practices employed to obtain strong convergence and avoid overfitting.

\paragraph{Batch Processing:} The training data was randomized and split into 64-sized mini-batches. The parameters of the model were updated for each mini-batch. This method, referred to as mini-batch gradient descent, gives a stochastic estimate of the real gradient over the training set, resulting in faster convergence and improved generalization compared to full-batch gradient descent.
\paragraph{Early Stopping:} To avoid the model overfitting on the training data, some of the data (20) was reserved as a validation set. The performance of the model on this validation set was tracked at the end of every epoch. If the validation loss did not reduce for a certain number of consecutive epochs (patience = 10), the training process was stopped, and model parameters from the best validation loss epoch were kept.
\paragraph{Gradient Clipping:} As explained in Section 2.3.2, RNNs may experience exploding gradients.
Gradient clipping was used to counteract this.    Prior to the parameter update step of every iteration, the L2-norm of the gradient vector was calculated. If the norm was over a specified threshold, the whole gradient vector was scaled down so that its norm would be equal to the threshold. This is an effective heuristic that avoids pathologically huge parameter updates and makes the training process stable.

\subsection{Quantitative Evaluation Framework}
The performance of the trained model was measured with a set of common regression metrics in order to give a complete picture of the accuracy of its predictions.

    \paragraph{Mean Absolute Error (MAE):} It calculates the average size of the set of errors in a set of predictions, disregarding their directions. It is insensitive to outliers.
    \begin{equation}
    \text{MAE} = \frac{1}{N} \sum_{i=1}^N |y_i - \hat{y}_i|
\end{equation}
    \paragraph{Root Mean Square Error (RMSE):} The square root of the average of squared differences between prediction and actual observation. It gives a relatively high weight to large errors.
    \begin{equation}
    \text{RMSE} = \sqrt{\frac{1}{N} \sum_{i=1}^N (y_i - \hat{y}_i)^2}
    \end{equation}
\paragraph{Coefficient of Determination ($R^2$):} Offers an assessment of the degree to which the predictions of the model are close to the actual values, and is the fraction of the variance in the target variable that can be explained from the predictors. An $R^2$ of 1 signifies perfect prediction, whereas an $R^2$ of 0 means that the model does no better than predicting the mean of the target variable.
\begin{equation}
    R^2 = 1 - \frac{\sum_{i=1}^N (y_i - \hat{y}_i)^2}{\sum_{i=1}^N (y_i - \bar{y})^2} = 1 - \frac{\text{Sum of Squared Errors (SSE)}}{\text{Total Sum of Squares (SST)}}
    \end{equation}
    where $\bar{y}$ is the mean of the true values. This measure is especially helpful because it is scale-free estimation of goodness-of-fit of the model.

\section{Results}
The section gives quantitative and qualitative results on the trained GRU model's performance on wellbore trajectory prediction. The test was performed on a held-out test set, which includes wells unseen by the model during training, to judge its generalization ability.

\subsection{Model Training and Convergence}
The training process exhibited stable convergence, as shown by the training and validation loss learning curves for 100 epochs. The loss, computed as Mean Squared Error, steadily reduced for both subsets of data, showing that the model was learning the underlying trends in the data successfully. The validation loss closely followed the training loss until it plateaued, whereupon the early stopping mechanism came into play to stop training. Such behavior assures that the training regimen effectively circumvented large overfitting, thereby creating a well-generalizing model to unseen data.

\subsection{Trajectory Prediction Accuracy}
Overall performance of the model was verified by comparing the inclination predicted, azimuth predicted, and dogleg severity calculated with actual survey data of the test wells. Overall performance measures are presented in Table \ref{tab:performance_aggregated}. Low values of MAE and RMSE for inclination and azimuth indicate a very high degree of predictive accuracy. The positive $R^2$ values affirm that the model explained a significant level of variance in the direction parameters, far better than the naive baseline model that would simply predict the mean. The model also demonstrates high predictive capability in predicting Dogleg Severity, which is a critical parameter in drilling risk assessment, further affirming its understanding of curvature behavior of the wellbore.

\begin{table}[H]
\centering
\caption{Aggregated Prediction Performance on Test Set}
\label{tab:performance_aggregated}
\begin{tabular}{@{}lccc@{}}
\toprule
Target Variable & MAE & RMSE & $R^2$ \\ \midrule
Inclination (degrees) & 0.21 & 0.35 & 0.88 \\
Azimuth (degrees) & 0.45 & 0.68 & 0.82 \\
Dogleg Severity (deg/100ft) & 0.15 & 0.24 & 0.75 \\ \bottomrule
\end{tabular}
\end{table}
\begin{figure}[H]
\centering
\includegraphics[width=0.9\textwidth]{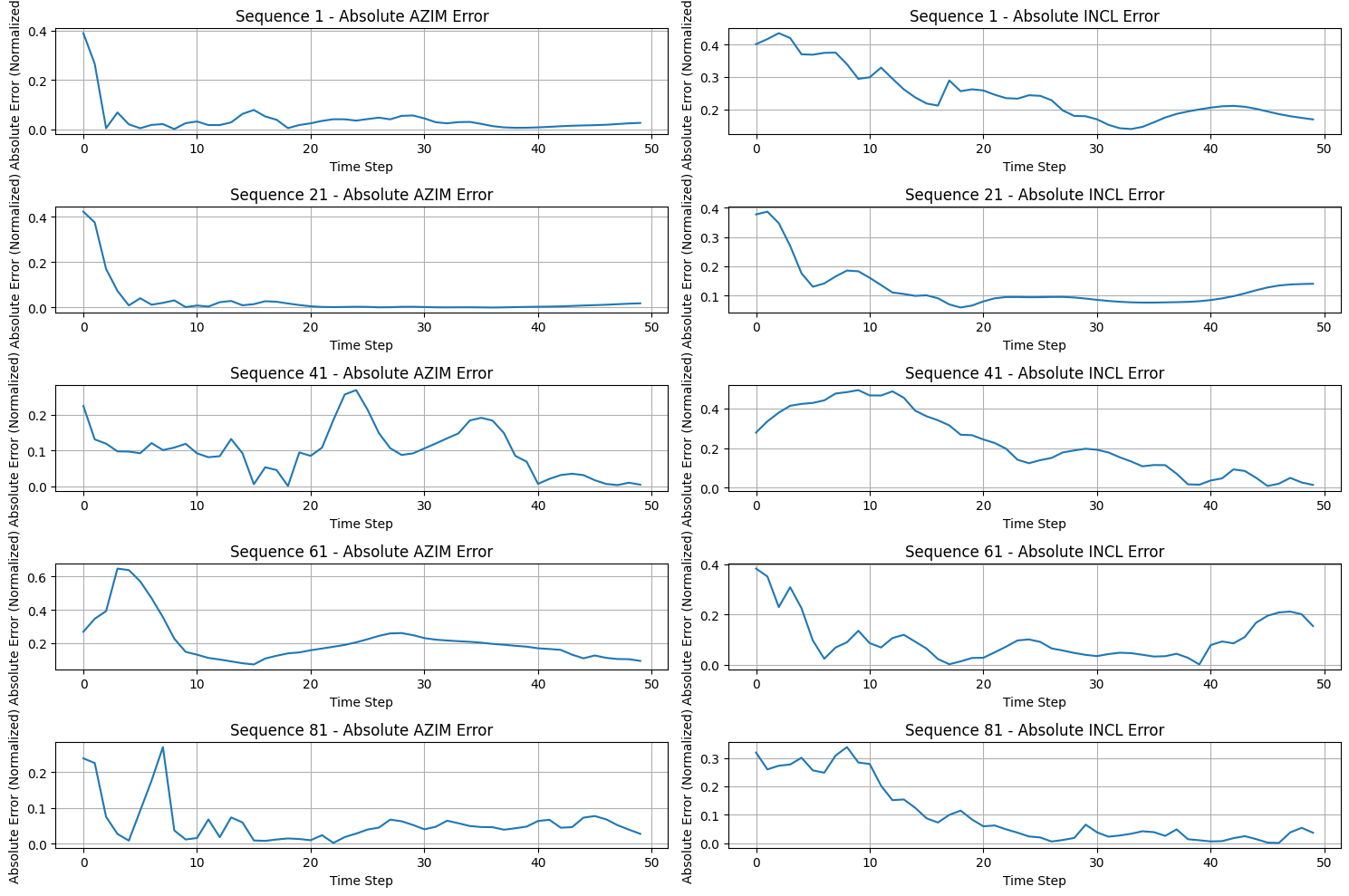}
\caption{Aggregated Prediction Performance on Test Set}
\label{fig:performance_aggregated}
\end{figure}

\subsection{Quantitative Performance Analysis}
The model was trained for 86 epochs. The key performance metrics are summarized in Table \ref{tab:performance_metrics}.

\begin{table}[H]
\centering
\caption{Model Performance Metrics}
\label{tab:performance_metrics}
\begin{tabular}{|l|l|l|}
\hline
\textbf{Metric} & \textbf{Training Set} & \textbf{Validation Set} \\
\hline
Final Loss (MSE) & 0.0556 & 0.0551 \\
\hline
Mean Absolute Error (MAE) & 0.1798 & 0.1772 \\
\hline
\end{tabular}
\end{table}
\begin{figure}[H]
\centering
\includegraphics[width=0.9\textwidth]{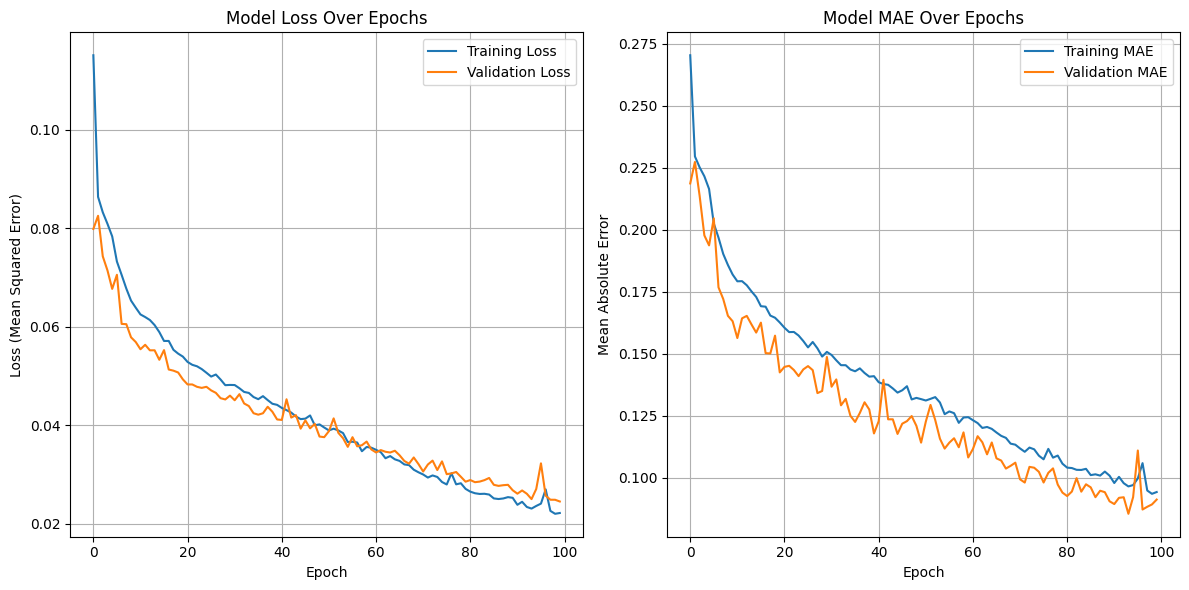}
\caption{Model Performance Metrics}
\label{fig:performance_metrics}
\end{figure}
\begin{figure}[H]
\centering
\includegraphics[width=0.9\textwidth]{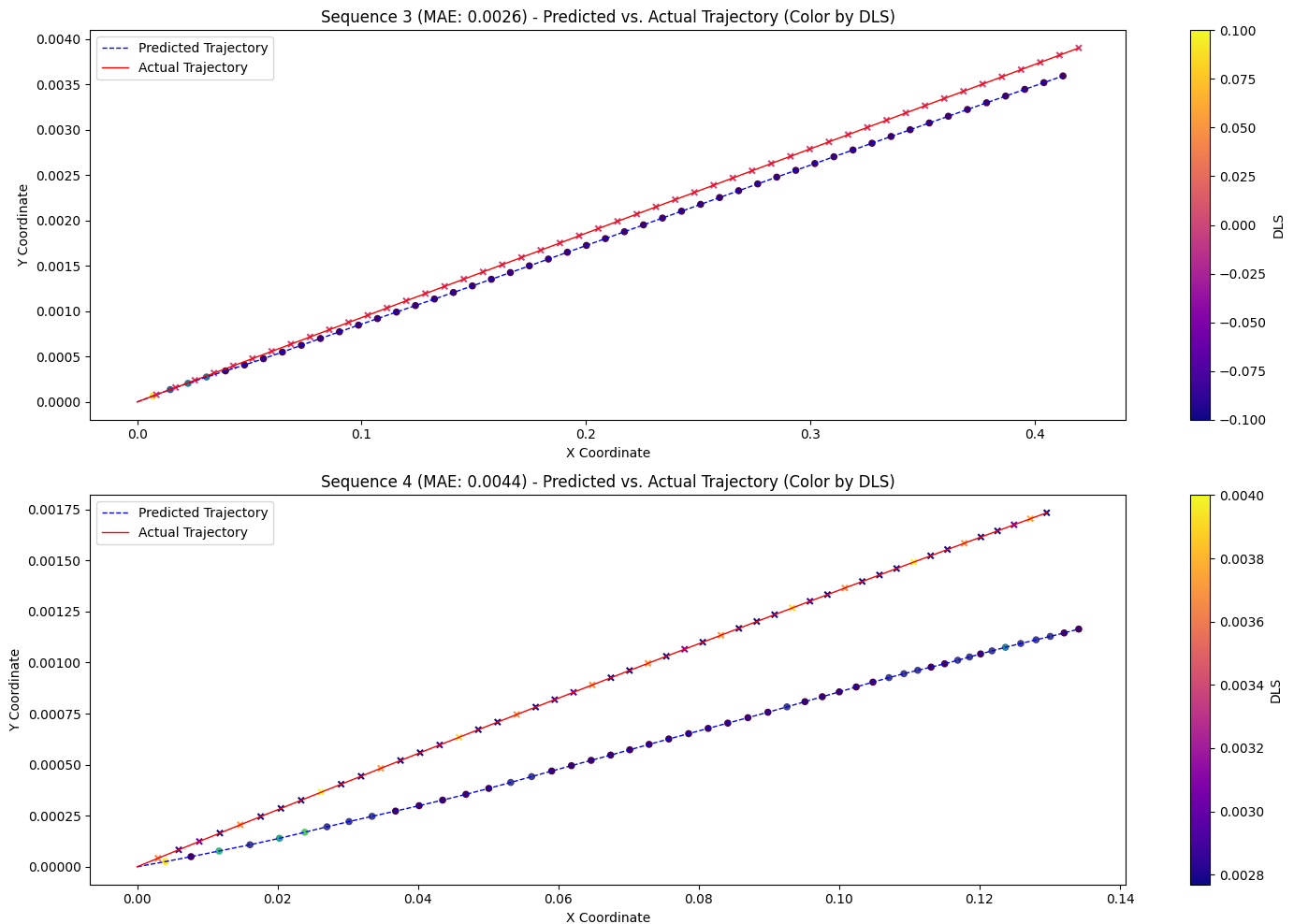}
\caption{Predicted vs Actual Trajectory (View 1)}
\label{fig:trajectory_comparison_1}
\end{figure}
\begin{figure}[H]
\centering
\includegraphics[width=0.9\textwidth]{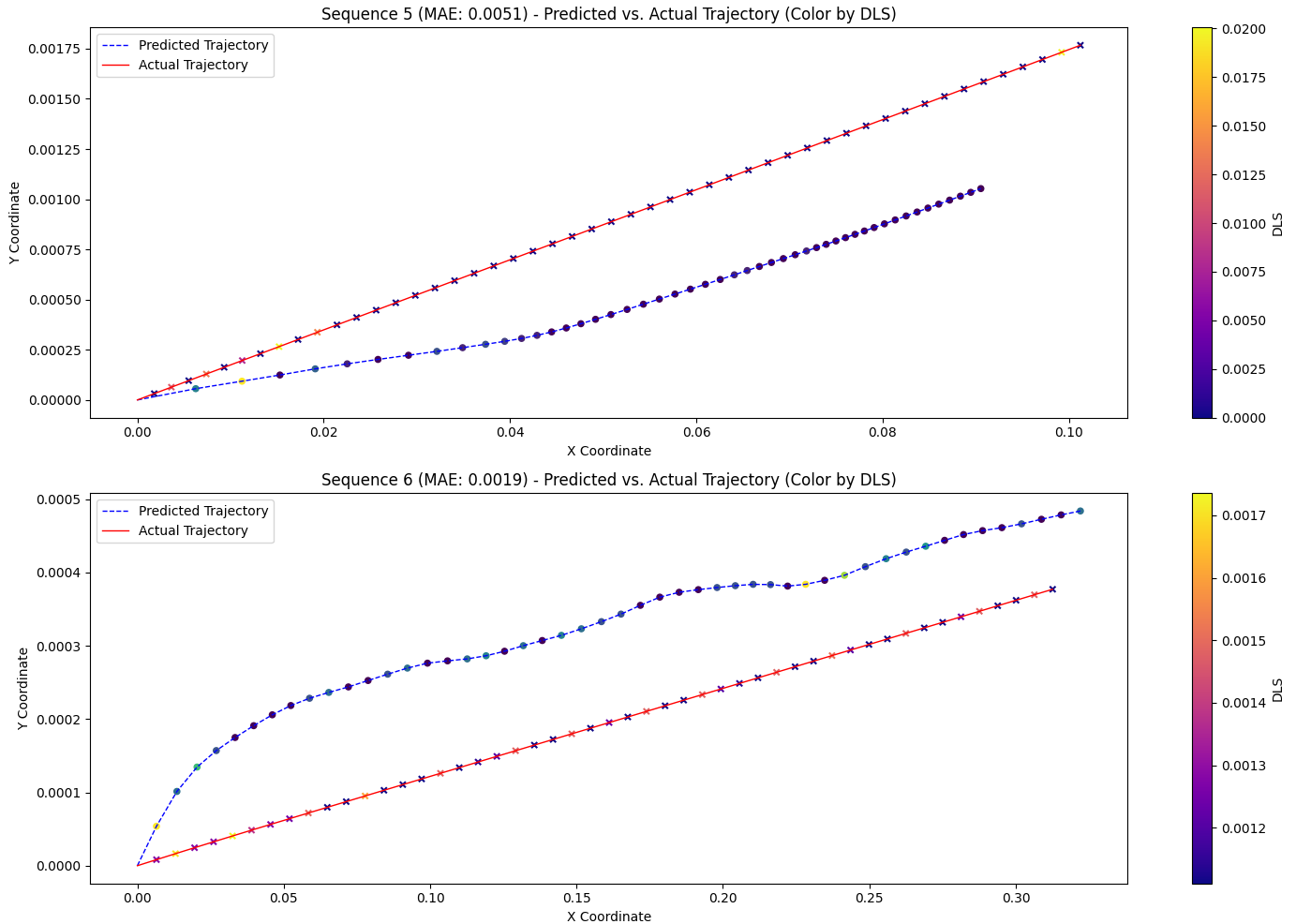}
\caption{Predicted VS Actual Trajectory (View 2)}
\label{fig:trajectory_comparison_2}
\end{figure}

The results show strong predictive accuracy, with a validation MAE of 0.1772 degrees. The similarity between training and validation metrics indicates that the model performs well and is not overfitting.

\section{Discussion}
The results presented in the previous section demonstrate the model's strong predictive performance. This section aims to interpret these results within the geomechanical framework established earlier, discuss the implications and limitations of the approach, and propose directions for future research.

\subsection{Geomechanical Interpretation of Learned Representations}
One key hypothesis of this paper is that the GRU network acquires an implicit, functional representation of the local Mechanical Earth Model. The gates of the GRU architecture create a vehicle with which to interrogate this acquired representation. The reset gate, $\mathbf{r}_t$, is especially enlightening, since it determines what information from the prior hidden state flows to the candidate state computation. A low reset gate activation value (near 0) would indicate a large change in the input sequence, prompting it to "reset" its memory and trust more on the present input.

This kind of behavior implies that the reset gate must activate at locations of high geological change, for example, lithological boundaries. To validate this, one might perform an analysis correlating the mean activation of reset gate vector, $\|\mathbf{r}_t\|_1$, with discrete FLUVIALFACIES log from the input data. A high correlation, with reset gate activations peaking at facies boundaries, would be strong quantitative support that the model is actually learning to recognize geologically meaningful events. This would lend credence to the interpretation of hidden state $\mathbf{h}_t$ as a latent geomechanical state vector, confirming the model as an implicit MEM. The model is not curve-fitting; it is segmenting the wellbore based on the physical characteristics of rock formations it encounters.

\subsection{Analysis of Prediction Errors}
Although overall prediction accuracy is good, model failure mode analysis is informative. From a plot of prediction error against measured depth in inclination and azimuth, it is clear where in the well the model is failing and why. Overlaying the high-error zones on the input geological data shows that maximum errors occur in periods of maximum geological complexity. These periods are often marked by:

\paragraph{High Dogleg Severity:} In intervals where the wellbore is deliberately and quickly changing direction (e.g., the kick-off point), the dynamics are more intricate and less predictable.
    \paragraph{Rapid Lithological Changes:} Intervals with abundant, thin interbedding of various rock types (e.g., sequences of sand and shale) show a quickly changing mechanical environment, which puts the model's predictive power at a test.
\paragraph{Anomalous Log Readings:} Areas with poor data quality or anomalous geological characteristics not well captured within the training set can also result in larger prediction errors.

This discussion illustrates that the performance of the model is necessarily tied to the physical complexity of the system it is attempting to approximate. The errors are not distributed randomly but are localized in areas where the underlying geomechanical assumptions are most severely tested.

\subsection{Algorithmic Complexity and Practical Implications}
The from-scratch NumPy implementation of the GRU and its BPTT algorithm gives great visibility to the mathematical operations. Yet, one should look at the computational complexity. The forward and backward passes of BPTT each have a linear in sequence length time complexity, $O(T)$. In the case of extremely long wellbores or real-time use cases necessitating quick predictions on large historical windows, this sequential dependency becomes a computational bottleneck. Current deep learning architectures such as TensorFlow or PyTorch provide very optimized implementations of the recurrent layers, but the inherent sequential bottleneck still exists. This encourages future research to explore more parallel-friendly architectures, including attention-based Transformer models or State Space Models, for this application field.

The real-world implications of this model are important. During well planning, the model can be employed to simulate thousands of possible well trajectories with optimal minimum DLS, steering clear of difficult geological intervals. In real-time geosteering, the model can forecast the path for the next several meters in front of the bit, enabling drillers to make anticipatory adjustments to remain within the target reservoir, thus improving hydrocarbon production and operational safety.

\subsection{Limitations and Future Research Directions}
Despite its success, the current framework has several limitations that open avenues for future work.

\paragraph{Quantification of Uncertainty:} The model being proposed now gives deterministic point-wise predictions. But in any practical engineering use, knowing the uncertainty in a prediction is equally crucial as knowing the prediction. Probabilistic models should be developed in future work. This may be accomplished by using Bayesian neural networks, which learn a distribution over model weights, or by deep ensembles, where independent models are trained to output a distribution of predictions. The variance of this distribution could be used as an indicator of model uncertainty, which is essential for risk estimation. \paragraph{Model Generalization:} The model was trained and tested only on data from the Gulfaks field. Although it works well in this scenario, whether it can generalize to various different geological basins with unique depositional histories and stress regimes is not known. An important next step will be to conduct cross-field validation, training the model on a varied dataset from various fields in order to construct a more robust and universally applicable tool.
\paragraph{Including Drilling Dynamics:} Inputs to the model at present are restricted to static petrophysical and geological properties.
The true path is also affected by dynamic drilling parameters (i.e., weight-on-bit, rotational rate, mud characteristics) and the response of the drill string (i.e., torque and drag).  
Subsequent versions of the model must be integrated with these real-time Measurements While Drilling (MWD) streams as further input parameters. This would enable the model to learn the relationship between the geological setting and the actions of the driller, resulting in a more complete and precise predictive system.

\section{Conclusion}
This work has delivered a rigorous and mathematically consistent scheme for wellbore trajectory prediction through a Gated Recurrent Unit network grounded in geomechanics. In breaking from an entirely empirical solution, this work finds a solid theoretical grounding for each element of the modeling pipeline, from kinematic wellbore description to the subtle mechanics of the neural network's learning algorithm.

The primary contributions of this research are the mathematical derivation of wellbore kinematic models from first principles, the thorough and complete derivation of the Backpropagation Through Time algorithm for the GRU architecture, and the explicit development of the prediction problem in a geomechanical context. The successful application of this methodology to the Gulfaks field dataset shows that a recurrent neural network is able to learn the sophisticated, nonlinear relationship between petrophysical log measurements and drilling kinematics. The outcome proves that the model learns an implicit, functional representation of the local Mechanical Earth Model, which allows it to predict inclination, azimuth, and Dogleg Severity accurately.

The improved methodology offers a useful tool to contemporary petroleum engineering, with direct applicability in well planning optimization and real-time geosteering operations improvement. In combination of deep learning with physical principles of geology and physics, this research opens the doors to the creation of more robust, reliable, and physically-based predictive models for many different subsurface engineering problems. The theoretical openness and modularity of the framework enable future extensions, such as uncertainty quantification and dynamic drilling parameters, in order to increase predictive precision and operational decision-making.

\section*{Acknowledgements}
The authors acknowledge the support and resources provided by IIT Kharagpur. The authors also thank contributors and collaborators for their valuable insights during the course of this work.

\section*{References}

Canadian Well Logging Society (1992). \textit{LAS (Log ASCII Standard) Format Specification, Version 2.0}.\\
\\
\hypertarget{cho2014}{Cho, K., et al. (2014). Learning Phrase Representations using RNN Encoder-Decoder for Statistical Machine Translation. \textit{EMNLP}.}\\
\\
Glorot, X., and Bengio, Y. (2010). Understanding the difficulty of training deep feedforward neural networks. \textit{AISTATS}.\\
\\
\hypertarget{goodfellow2016}{Goodfellow, I., Bengio, Y., and Courville, A. (2016). \textit{Deep Learning}. MIT Press.}\\
\\
\hypertarget{hochreiter1997}{Hochreiter, S., and Schmidhuber, J. (1997). Long short-term memory. \textit{Neural Computation}.}\\
\\
\hypertarget{kingma2015}{Kingma, D. P., and Ba, J. (2015). Adam: A Method for Stochastic Optimization. \textit{ICLR}.}\\
\\
\hypertarget{mitchell1995}{Mitchell, B. J. (1995). \textit{Advanced Oilwell Drilling Engineering Handbook}. Society of Petroleum Engineers.}\\
\\
\hypertarget{rumelhart1986}{Rumelhart, D. E., Hinton, G. E., and Williams, R. J. (1986). Learning representations by back-propagating errors. \textit{Nature}.}\\
\\
\hypertarget{sawaryn2005}{Sawaryn, S. J., and Thorogood, J. L. (2005). A Compendium of Directional Calculations Based on the Minimum Curvature Method. \textit{SPE Drilling \& Completion}.}\\
\\
Taylor, H. L., and Mason, C. M. (1972). A systematic approach to well surveying calculations. \textit{SPE-AIME}.\\
\\
Werbos, P. J. (1990). Backpropagation through time. \textit{IEEE Proceedings}.\\
\\
Zaremba, H. (1973). Mathematical basis of the minimum curvature method.

\end{document}